# Coordination-Induced Tuning of Ligand-Centered Red Emission in a *cis*-$[Cd(Tz)_2(py)_2]$ Complex for Light-Emitting Diodes

Samara M. da Silva[1‡], R. F. Silva[6], A. Nonato[2,3‡*], Paulo Villis[4], Rodrigo S. Corrêa[5], L. C. Gómez-Aguirre[2], C.W.A. Paschoal[6], Pedro I. S. Maia[7], Benedicto A. V. Lima[8,*]

[1]*Coordenação de Ciências Naturais - Química, Universidade Federal do Maranhão, Centro de Ciências de Grajaú- CCGR, 65.940-000, Grajaú, MA, Brazil.*

[2]*Department of Thermal Energy Storage, Iberian Centre for Research in Energy Storage CIIAE, 10003 Caceres, Spain*

[3]*Programa de Pós-graduação em Física, Universidade Federal do Maranhão, Campus Universitário do Bacanga, São Luís, Maranhão 65080-805, Brazil*

[4]*Laboratory of Electrochemistry and Biotechnology, Ceuma University, São Luís, postal code 65.065-470, MA, Brazil.*

[5]*Departamento de Química, Instituto de Ciências Exatas e Biológicas (ICEB), Federal University of Ouro Preto, Morro do Cruzeiro Campus, postal code 35.400-000, Ouro Preto, MG, Brazil.*

[6]*Departamento de Física, Universidade Federal do Ceará, Campus do Pici, 65455-900, Fortaleza - CE, Brazil*

[6]*Departamento de Química, Instituto de Ciências Exatas e Biológicas (ICEB), Universidade Federal de Ouro Preto, 35400-000 Ouro Preto-MG, Brazil*

[7]*Núcleo de Desenvolvimento de Compostos Bioativos (NDCBio), Universidade Federal do Triângulo Mineiro, Av. Dr. Randolfo Borges 1400, Uberaba, Minas Gerais, Brazil*

[8]*Medicinal Chemistry and Biotechnology Research Group, Coordenação de Ciências Naturais - Química, Universidade Federal do Maranhão, Centro de Ciências de São Bernardo - CCSB, 65550-000, São Bernardo, MA, Brazil*

[*]Corresponding author. Tel: +34 610281478

E-mail address: ariel.almeida@ciiae.org; benedicto.lima@ufma.br

‡ These authors contributed equally to this work.

# ABSTRACT

Organic–inorganic complexes have emerged as highly promising materials for light-emitting applications. In this work, we report a new organometallic complex, *cis*-[Cd(1,3-bis(2-methoxy-4-nitrophenyl)triazene)$_2$(pyridine)$_2$] (*cis*-[Cd(Tz)$_2$(py)$_2$]), featuring a distorted octahedral Cd(II) coordination environment. Vibrational spectroscopy (IR and Raman) reveals pronounced spectral changes upon coordination, particularly in the Raman response of the triazene moiety, evidencing strong electronic and structural perturbation induced by complexation with the Cd(II) center. Hirshfeld surface analysis reveals that the crystal packing is predominantly governed by H···H, O···H/H···O, and C···H/H···C interactions, while π–π stacking interactions, although structurally evident, contribute only modestly to the intermolecular packing. The solid-state UV–Vis absorption spectrum exhibits a broad profile ranging from ~700 to 200 nm, indicating multiple electronic transitions and highlighting its potential for optoelectronic applications. Tauc analysis reveals a direct optical band gap of 1.83 eV, suggesting semiconductor-like behavior. Photoluminescence studies evidence that the emission arises predominantly from ligand-centered (LC) transitions (π→π and n→π), consistent with the $d^{10}$ electronic configuration of Cd(II), which prevents metal-centered and charge-transfer transitions (MLCT/LMCT). Upon coordination, a pronounced modification of the emission profile is observed, including band broadening and enhanced red emission (500–850 nm), associated with changes in excited-state dynamics, increased structural rigidity, and modulation of non-radiative relaxation pathways. The CIE chromaticity coordinates shift from (0.5521, 0.4417) for Tz to (0.5229, 0.4597) for the complex, both located in the warm emission region. These findings demonstrate the potential of this system for application in red-emitting light-emitting diodes.

# 1. Introduction

Over the past decades, metal–organic and organic–inorganic hybrid materials have attracted significant attention due to their potential applications in light-emitting devices, sensors, displays, and other optoelectronic technologies [1–8]. These systems provide versatile platforms for luminescence generation, especially when combining ‘transition metal ions with organic ligands containing aromatic or π-conjugated systems [9–11]. Moreover, the interplay between inorganic and organic components often leads to materials that not only retain the intrinsic properties of each constituent but also exhibit new or enhanced photophysical behaviors arising from cooperative interactions [12–14].

Among transition metal coordination complexes (TMCC), systems containing divalent ions with a $d^{10}$ electronic configuration have been extensively investigated due to their distinct structural and photophysical characteristics. These complexes are typically characterized by high kinetic lability, allowing ligand substitution processes to occur more readily compared to other transition metal systems [15,16]. As a consequence, complexes based on Zn(II), Cd(II), and Hg(II) are frequently obtained, enabling the exploration of a wide range of ligands in coordination chemistry. In particular, the Cd(II) ion has emerged as an attractive center owing to its flexible coordination environment and rich structural chemistry. Cadmium complexes can adopt a wide range of coordination numbers, typically varying from four to eight, resulting in diverse geometries such as tetrahedral, square planar, trigonal bipyramidal, octahedral, and more complex polyhedral [17–20]. This structural diversity is mainly attributed to the relatively large ionic radius and the $d^{10}$ electronic configuration of Cd(II), which leads to negligible crystal field stabilization energy. Consequently, the final structure is predominantly governed by metal–ligand interactions and steric factors, allowing for effective tuning of material properties through ligand design. As a result, numerous studies have reported Cd(II) coordination polymers and multinuclear complexes constructed from a wide variety of ligands. [12,18,19,21–25]

In addition to their structural versatility, Cd(II)-based complexes are particularly appealing for photoluminescent applications. Due to the fully filled d orbitals, metal-centered and charge-transfer transitions (MLCT/LMCT) are generally suppressed, and the emission typically arises from ligand-centered (LC) excited states, such as $\pi\rightarrow\pi^*$ and $n\rightarrow\pi^*$ transitions [26]. These characteristics make cadmium complexes excellent model systems for investigating intraligand emission processes and their modulation by coordination. Furthermore, intermolecular interactions in the solid state, including hydrogen bonding, $\pi$–$\pi$ stacking, and other weak contacts, can significantly influence the rigidity of the crystal lattice and, consequently, the photophysical response of the material.

Luminescence in such systems arises from the relaxation of electronically excited states to lower-energy states. In organic ligands, the most common electronic transitions involved in this process are $\pi\rightarrow\pi^*$ and $n\rightarrow\pi^*$. While $\pi\rightarrow\pi^*$ transitions are typically associated with strong emission, $n\rightarrow\pi^*$ transitions often involve lower energies and can be less efficient due to their partially forbidden nature [27]. Upon coordination to metal centers, however, these transitions can be altered, leading to changes in emission intensity, spectral profile, and overall luminescence efficiency. These effects are frequently associated with energy transfer processes and the activation of non-radiative relaxation pathways, which can result in fluorescence quenching or redistribution of emission intensity [10,28].

In this context, the design of cadmium-based complexes incorporating π-conjugated ligands represents a promising strategy for developing new luminescent materials with tunable optical properties. Herein, we report the synthesis and characterization of a new organometallic complex, *cis*-[Cd(1,3-bis(2-methoxy-4-nitrophenyl)triazene)$_2$(pyridine)$_2$ (*cis*-[Cd(Tz)$_2$(py)$_2$]), featuring a distorted octahedral coordination environment. The structural, vibrational, and optical properties of the complex were investigated in detail, with particular emphasis on its photoluminescent behavior. The results demonstrate that coordination significantly influences the electronic structure of the ligand, leading to enhanced emission in the red region and highlighting the potential of this system for application in light-emitting devices.

## 2. Experimental procedures

### Materials

All chemicals used were of analytical reagent grade and all solvents used in this study were used without purification (Sigma-Aldrich).

### Synthesis

1,3-bis(2-methoxy-4-nitrophenyl) triazene was synthesized as previously described [35]. For the synthesis of *cis*-[Cd(1,3-bis(2-methoxy-4-nitrophenyl) triazene)2(pyridine)2] 138.91 mg (0.4 mmol) of 1,3-bis(2-methoxy-4-nitrophenyl) triazene (Tz) was dissolved in 75 ml of acetone in a round-bottom flask and 240 μL of a KOH 1,78 mol/L solution in methanol was added. The orange solution turned purple immediately. Then, 40.26 mg (0.2 mmol) of $CdCl_2 \cdot H_2O$ was added to the mixture. The purple solution turned red. The system was left under stirring for 1 h and 120 μL (0.435 mmol) of pyridine (py) was added. The system was left to react under stirring for 1.5 h. The system was filtered off, the volume was reduced under vacuum to approximately 5 mL, precipitated and washed with water repeatedly and dried. The solid obtained was washed with ether, dissolved in a mixture acetone/pyridine (10:1) solvent and allowed to crystallize for one week.

**X-ray Single Crystal Structure Determination**

The crystal was grown by slow evaporation of a acetone/methanol solution. It was mounted on an Enraf-Nonius Kappa-CCD diffractometerwith graphite monochromated Mo Kα radiation (λ = 0.71073 Å). The final unit cell parameters were based on all reflections. Data collection was performed using the COLLECT program [29], integration and scaling of reflections were performed with the HKL DENZO-SCALEPACK software [30]. Absorption correction was performed using the Gaussian method [31]. The structure was solved by direct methods with SHELXS-97 and its model was refined by full matrix at least squares in F2 using SHELXL-97 [32]. All hydrogen atoms were positioned stereochemically and refined with a rigid refinement model. The ORTEP view was prepared using ORTEP-3 for Windows [33]. The hydrogen atoms in the aromatic rings were refined isotropically, each with a thermal parameter 20% larger than the equivalent isotropic shift parameter of the atom to which it is bonded. CCDC 2531833 contains the supplementary crystallographic data for this paper. These data can be obtained free of charge via the joint Cambridge Crystallographic Data Centre (CCDC) and Fachinformationszentrum Karlsruhe.

**Infrared Spectroscopy**

The attenuated total reflectance (ATR) spectrum was obtained using a Bruker Vertex 70 v Fourier transform spectrometer. Data in the mid- and far-infrared regions were collected using a Globar source, with 128 scans and a spectral resolution of 2 $cm^{-1}$ in the 70-4000 $cm^{-1}$ range. Signal detection was performed using DLaTGS pyroelectric detectors. The ATR spectra were converted into absorbance spectra using the OPUS software.

**Raman Spectroscopy**

The Raman spectra at room temperature were recorded in the 80-3000 $cm^{-1}$ range with a 785 nm diode laser as the excitation source using a Horiba LabRAM HR spectrometer equipped with a charge-coupled device (CCD) cooled with liquid nitrogen. For *cis*-[$Cd(Tz)_2(py)_2$] complex, the measurements were performed with 5 accumulations and an exposure time of 10 seconds. For sample Tz, 10 accumulations of 20 seconds were used. The spectra were deconvolved into a sum of Lorentzian functions using the Fityk software. [34]

**Photoluminescence Spectroscopy**

Photoluminescence (PL) emission spectra at room temperature were obtained in the range of 470 to 850 nm using the T64000 spectrometer (Horiba), equipped with an Olympus microscope, a 20x objective lens, a charge-coupled device (CCD) cooled with liquid nitrogen, and a 457 nm argon ion laser as the excitation source. The spectrum of the complex was obtained with 3 accumulations and 5 seconds of exposure, while in the measurement of the Tz sample, 3 accumulations of 1 second were performed.

**Diffuse Reflectance Spectroscopy (DRS)**

The room temperature reflectance spectra of the two powder samples were collected using a Cary 7000 spectrophotometer (Agilent Technologies), equipped with an integrating sphere, in the range between 200 and 800 nm, with an average exposure time of 1 s and a scan rate of 60 nm/min. The spectra were converted to absorption spectra using the Kubelka-Munk function [35]. Bandgap energies were estimated using the Tauc method [36].

**Computational approach**

Three-dimensional Hirshfeld surfaces were generated using the CrystalExplorer 17.5 software package. [37] The surfaces were mapped over the normalized contact distance ($d_{norm}$), which is defined in terms of the distances from the surface to the nearest internal ($d_i$) and external ($d_e$) atoms, relative to their van der Waals radii. The $d_{norm}$ parameter is expressed as:

$$d_{norm} = \frac{d_i - r_i^{vdW}}{r_i^{vdW}} + \frac{d_e - r_e^{vdW}}{r_e^{vdW}} \qquad (1)$$

where $r_i^{vdW}$ and $r_e^{vdW}$ correspond to the van der Waals radii of the atoms located inside and outside the surface, respectively. Mapping the Hirshfeld surface over $d_{norm}$ allows the visualization of close intermolecular contacts, which appear as highlighted regions on the surface. [38] In addition, two-dimensional (2D) fingerprint plots were generated based on $d_i$ and $d_e$ distances. These plots provide a comprehensive representation of all intermolecular interactions and enable the quantitative assessment of different contact types through their percentage contributions. [39]

# 3.Results and discussions

*3.1 Chemistry*

Complex *cis*-[Cd(Tz)$_2$(py)$_2$] was synthesized according to Scheme 1. Dissolution of TzH in acetone yields a clear orange solution that turns into a deep purple color after addition of KOH solution evidencing the deprotonation of TzH. Addition of $CdCl_2 \cdot H_2O$ leads to a formation of a dark red solution in a few seconds due to the coordination of Tz to cadmium. Posterior addition of pyridine does not change the solution color significantly. Slow evaporation of solvents of a saturated solution of *cis*-[Cd(Tz)$_2$(py)$_2$] in acetone/pyridine allowed the formation of crystals which were used for chemical analysis.

Scheme 1. Synthesis reaction of the complex *cis*-[Cd(Tz)$_2$(py)$_2$].

*3.2. Room temperature Raman spectroscopy and FTIR*

The IR and Raman spectra of the *cis*-[Cd(Tz)$_2$(py)$_2$] complex recorded at room temperature are shown in Figures 1 and 2, respectively. According to the factor group analysis, considering the symmetries of the sites occupied in the crystal structure within the $P2_1/n$ (no. 14, Z = 4) space group, 1161 optical vibrational modes are expected. Among these modes, 582 ($291A_g \oplus 291B_g$) are Raman-active, and 579 ($290A_u \oplus 289B_u$) are infrared-active. In this centrosymmetric space group ($C_{2h}$, 2/m), which contains an inversion center, the mutual exclusion rule applies. Therefore, Raman-active modes (g) and infrared-active modes (u) are mutually exclusive, and no vibrational mode is simultaneously active in both Raman and infrared spectra.

For comparison purposes, the vibrational spectra of the free ligand (Tz), including both IR and Raman data, are provided in the Supplementary Material (Figures S1–S2). As observed, the Raman spectrum undergoes more significant changes upon coordination than the IR spectrum, which remains comparatively unchanged. In general, the Raman spectra are more selective than the IR spectra, and this is consistent with the $\pi$ structure of the derivative [40,41].

The vibrational modes of the *cis*-[Cd(Tz)$_2$(py)$_2$] complex can be assigned based on the vibrational features of triazine units previously reported in the literature [42–46]. Considering this approach, the spectral bands can be classified according to distinct regions. In the high-frequency region (~3000 cm$^{-1}$), the bands are attributed to C–H stretching vibrations. In the intermediate region (1500–1600 cm$^{-1}$), the bands correspond to ring stretching modes, analogous to those observed in benzene. The region between 1200 and 1400 cm$^{-1}$ is associated with C–H rocking vibrations, while at lower wavenumbers (below 1000 cm$^{-1}$), additional ring deformation modes are observed. A complete assignment of the vibrational modes is also given in the Supplementary Material (Table S1).

A comparative analysis of the free (Tz) and coordinated ligand reveals significant spectral changes upon complexation (see Fig. S1). In particular, the intense band at 1346 cm$^{-1}$, assigned to C–H rocking vibrations, shows a frequency shift, indicating changes in the ligand environment. This band, appearing at 1334 cm$^{-1}$ in the free ligand, undergoes a blue shift to 1346 cm$^{-1}$ upon coordination and is observed in both IR and Raman spectra. A similar behavior is observed for the band around 921 cm$^{-1}$, assigned to asymmetric N–C and C–C stretching coupled with $CH_2$ rocking vibrations, which is characteristic of the triazine moiety [44,47]. Upon coordination, this band undergoes significant changes in both position and intensity, reflecting the strong influence of Cd(II) coordination on the electronic structure and vibrational dynamics of the triazine unit. Additionally, the low-frequency region (200–400 cm$^{-1}$), attributed to ring deformation modes, shows pronounced changes upon coordination, characterized by a blue shift of the vibrational modes.

The spectral changes observed in *cis*-[Cd(Tz)$_2$(py)$_2$] are further evidenced by the disappearance of prominent bands at 1258, 1406, and 1435 cm$^{-1}$, originally associated with the free Tz ligand. Nevertheless, several vibrational features remain preserved, indicating that key structural elements of the triazine moiety are retained upon coordination. Overall, the combined IR and Raman analyses support the proposed coordination environment of *cis*-[Cd(Tz)$_2$(py)$_2$].

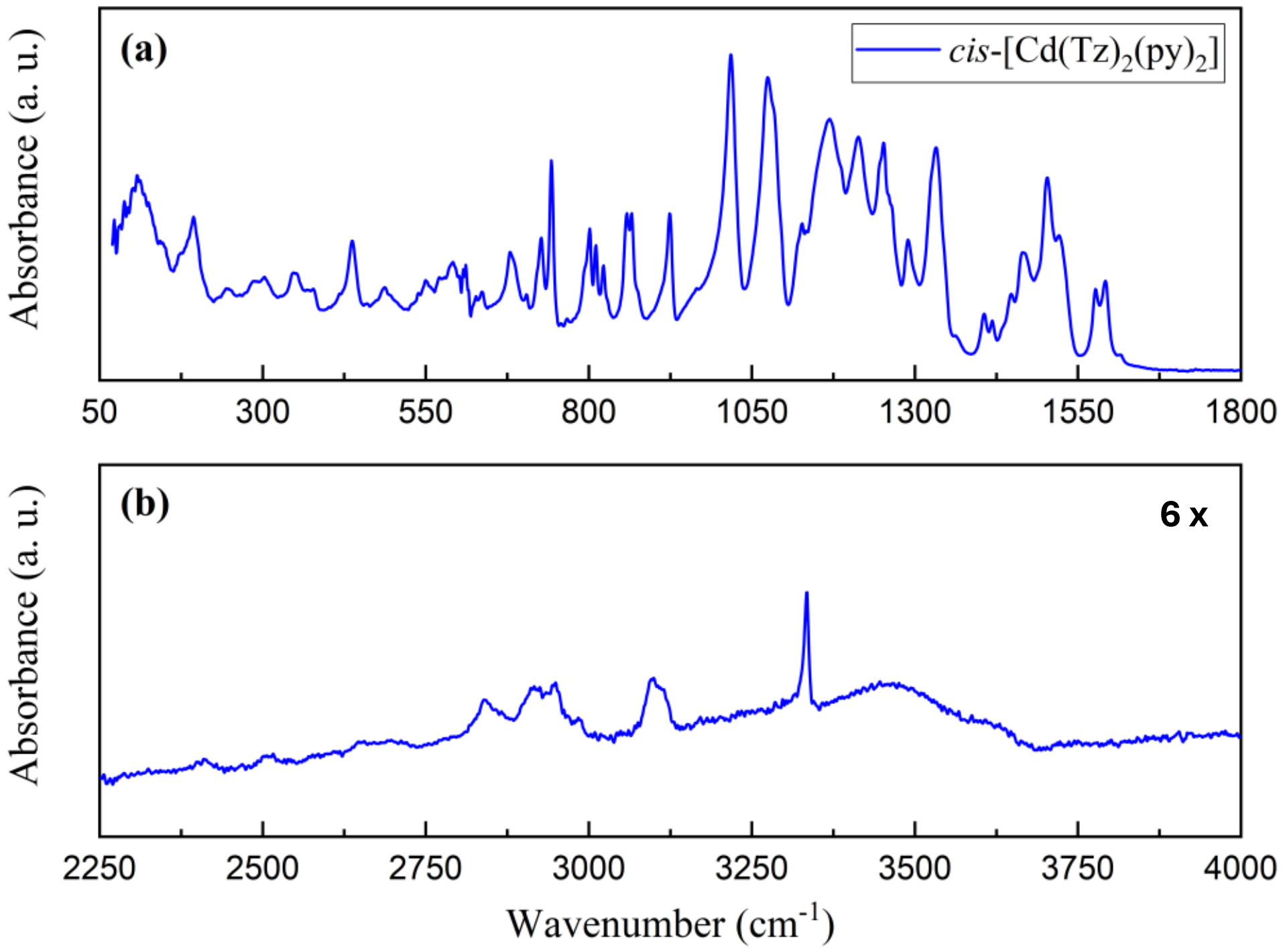

**Figure 1**. ATR spectrum of *cis*-[Cd(Tz)$_2$(py)$_2$] at 300 K in the range 70 to 4000 cm$^{-1}$.

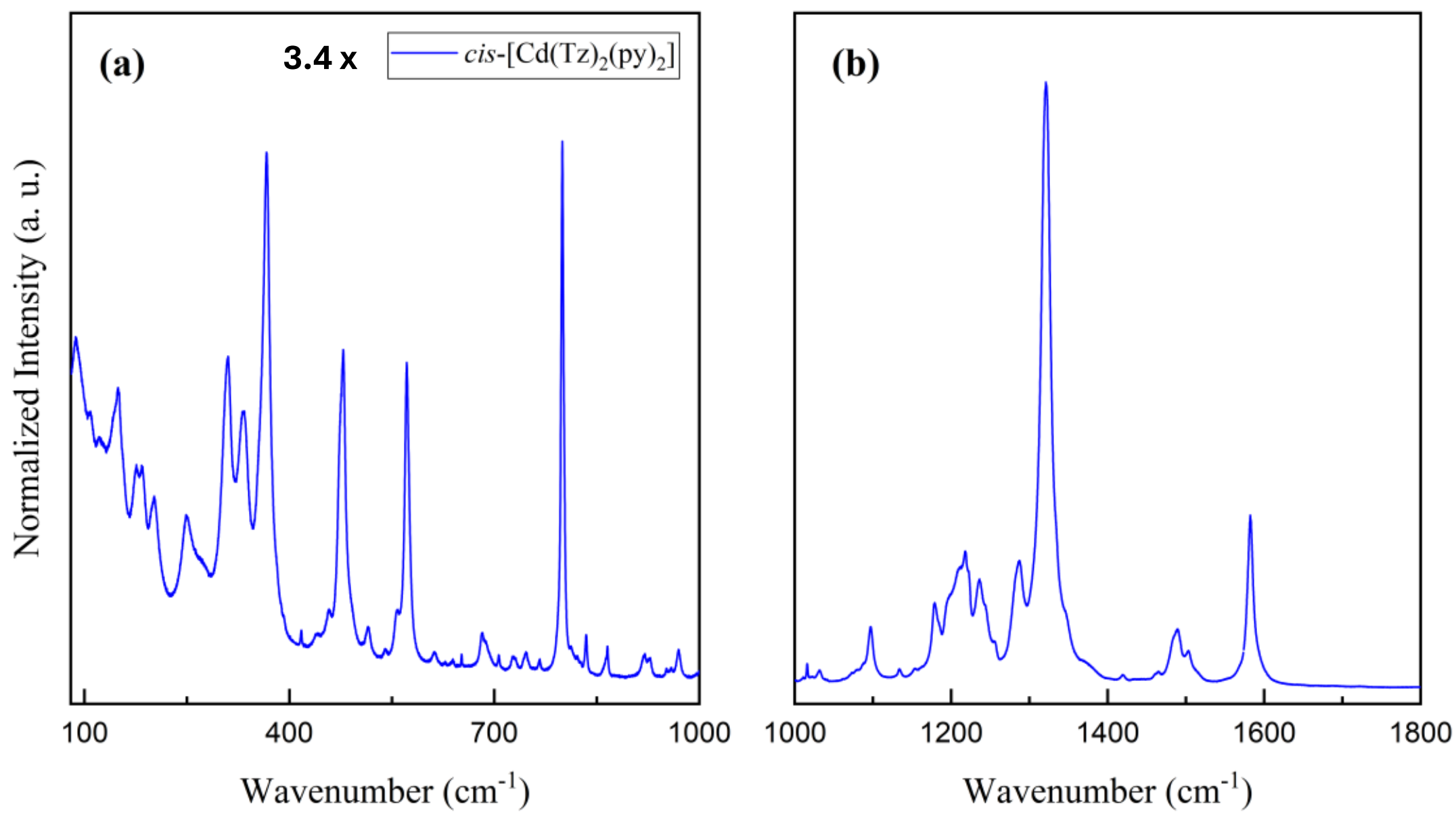

**Figure 2**. Raman spectrum of *cis*-[Cd(Tz)$_2$(py)$_2$] at 300 K with normalized intensity in the range 80 to 1000 cm$^{-1}$.

*3.3. X-ray crystallography*

The geometry and mode of coordination of *cis*-[Cd(Tz)$_2$(py)$_2$] could be determined by the X-ray crystallographic analysis. It revealed a distorted octahedral structure around the central atom, with both Tz ligands coordinated in a bidentate way. The asymmetric unit of the complex contains one Cd(II) atom, two pyridines, and two Tz anionic ligands. X-ray structure of the compound is represented in Figure 3. The crystallographic parameters and details regarding data collection and refinement of the complex are listed in Table 1 and selected bond distances and angles are presented in Table 2.

The distorted octahedral geometry presented N(1)-Cd1-N(21), N(1)-Cd1-N(12), N(12)-Cd1-N(21) and N(11)-Cd1-N(12) angles equal to 99.16(14), 131.31(15), 100.61(16) and 91.51(17)$^{\circ}$, respectively (Fig. 3). In an ideal octahedral structure these angles should all measure 90º. The angle between N(11)-Cd1-N(12) and N(1)-Cd1-N(21) planes is equal to 62.67º, in a perfect octahedron these planes would be coplanar.

The distances Cd1-N(1) = 2.218(4) Å, Cd1-N(21) = 2.290(4) Å are close to the ones reported for the structurally similar complex *trans*-Bis[1,3-bis(2-methoxyphenyl)-triazenido]dimethanolcadmium(II), which presents Cd-N bond length equal to 2.188 (4) and 2.230 (3) Å [48]; but are shorter than those observed for *cis*-Bis[1,3-bis(2-fluorophenyl)triazenido-κ$^2$N$^1$,N$^3$]bis(pyridine-κN)-cadmium(II), which present Cd-N distance around 2.37 Å [49]. The distances of the N-N bonds in the free ligand measure 1.294(3) and 1.314(2) Å [50], in the compound *cis*-$[Cd(Tz)_2(py)_2]$ these distances are 1.305(6) Å for the N21-N22 bond and 1.292(6) Å for the N22-N23 bond, presenting no significant differences.

Small contraction of N-N-N angle is caused by coordination of Tz to cadmium, in the free ligand N-N-N = 112.1(1)º, whereas in the complex this angle measures 110.8(4)º, evidencing coordination of Tz to cadmium in a bidentate way. *Trans* Cd(1)-N(3) and Cd(1)-N(23) bond lengths are significantly longer steric effect caused by nitrogen atoms (Cd1-N3 = 2.768(9) and Cd1-N23 = 2.682(4) Å), presenting N(3)-Cd(1)-N(23) angle equal to 176.9º.

Intramolecular π–π stacking interactions were observed between phenyl rings, with a centroid–centroid distance of 3.69 Å and an interplanar angle of 13° (Fig. 4). In the crystal structure, intermolecular π–π stacking interactions are also present between phenyl rings of Tz moieties in adjacent molecules. The phenyl rings within each Tz unit are nearly coplanar, with an interplanar angle of 1.2°, while those involved in intermolecular interactions are almost parallel, with a dihedral angle of only 0.6°. The corresponding intermolecular centroid–centroid distance is 3.68 Å (Fig. 2).

These features indicate that π–π interactions contribute to the organization of the crystal packing. In addition, other intermolecular interactions are observed, including σ–lone pair contacts between nitrite oxygen atoms and methylene groups (3.086 Å), as well as Cπ···Cπ interactions between phenyl and pyridine rings (3.383 Å).

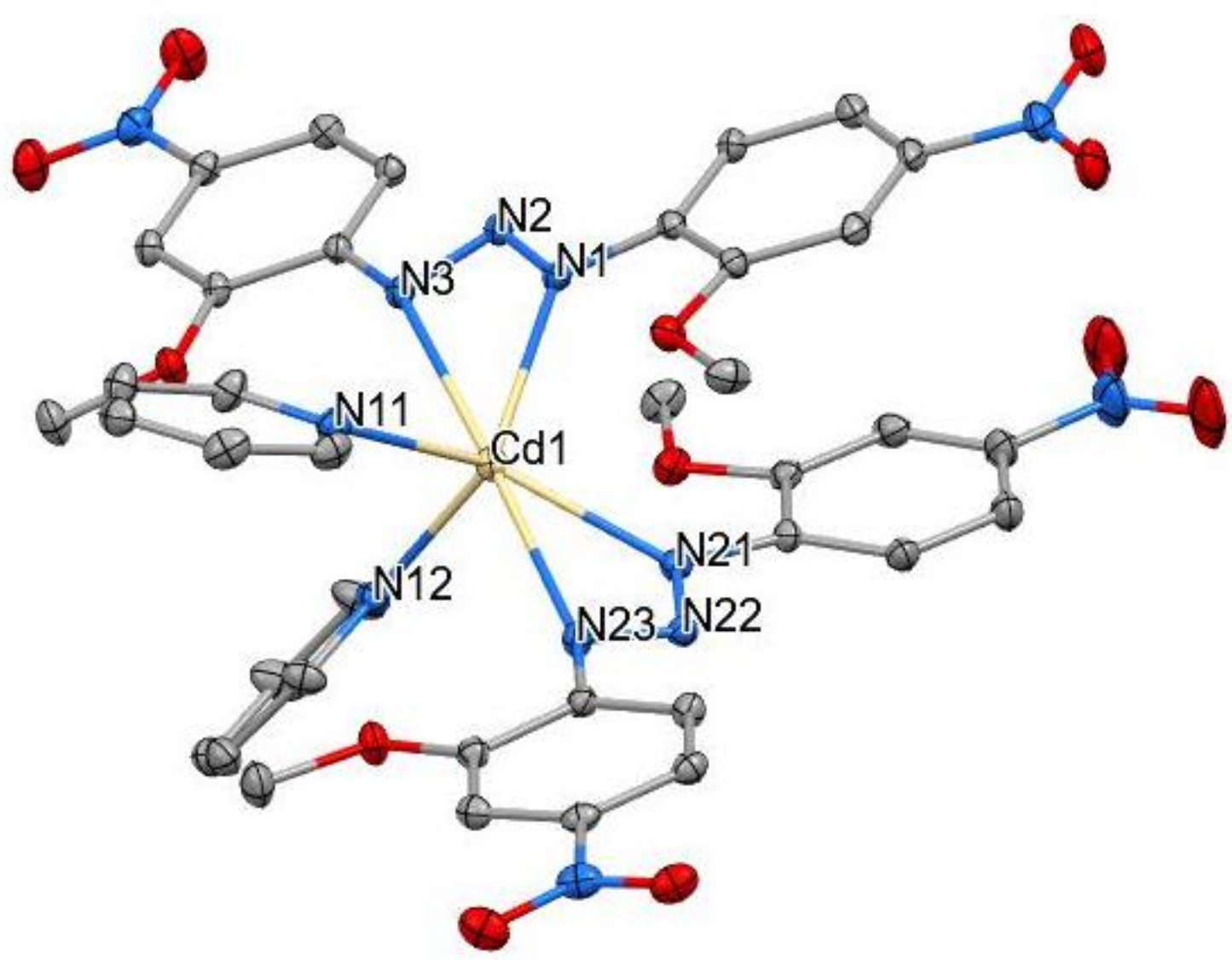

Figure 3. X-ray structure of *cis*-[Cd(Tz)$_2$(py)$_2$], for clarity only selected atoms are labeled and ellipsoids are drawn at the 30% probability level.

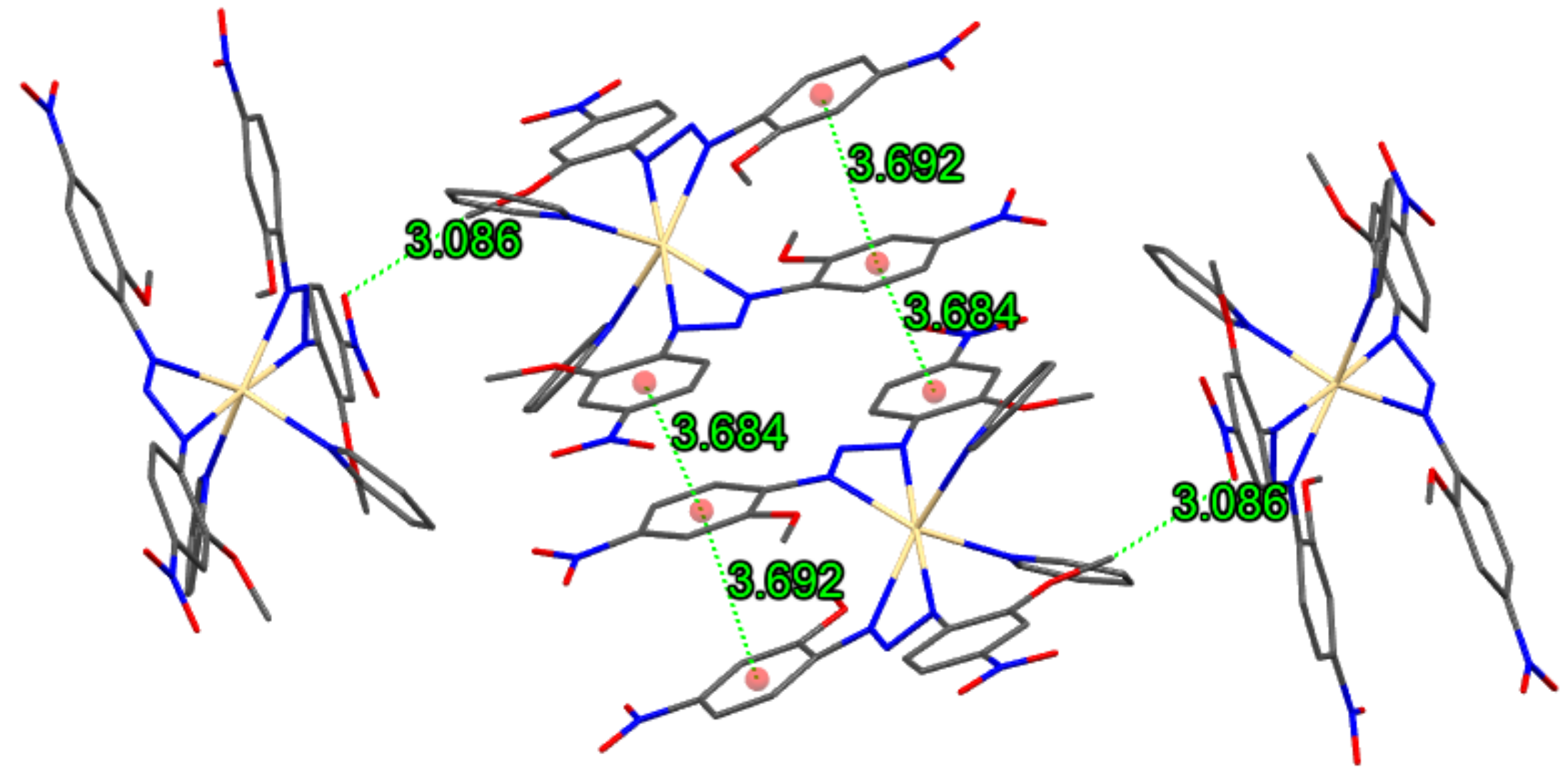

Figure 4. Intermolecular and intramolecular π-π stacking; Cσ-lone pair and Cπ···Cπ interaction in *cis*-[Cd(Tz)$_2$(py)$_2$] crystal packing.

Table 1. Crystal data and structure refinement for *cis*-[$Cd(Tz)_2(py)_2$].

| | | |
|---|---|---|
| Empirical formula | $C_{38}H_{34}CdN_{12}O_{12}$ | |
| Formula weight | 963.17 | |
| Temperature | 273(2) K | |
| Wavelength | 1.54178 Å | |
| Crystal system | Monoclinic | |
| Space group | P 21/n | |
| Unit cell dimensions | a = 11.3351(17) Å | a= 90°. |
| | b = 27.801(4) Å | b= 107.827(7)°. |
| | c = 13.920(2) Å | g = 90°. |
| Volume | 4176.0(12)$Å^3$ | |
| Z | 4 | |
| Density (calculated) | 1.532 $Mg/m^3$ | |
| Absorption coefficient | 4.853 $mm^{-1}$ | |
| F(000) | 1960 | |
| Crystal size | 0.729 x 0.324 x 0.130 $mm^3$ | |
| Theta range for data collection | 3.179 to 74.931°. | |
| Index ranges | -14<=h<=13, -34<=k<=34, -17<=l<=17 | |
| Reflections collected | 58861 | |
| Independent reflections | 8481 [R(int) = 0.0809] | |
| Completeness to theta = 67.679° | 99.5 % | |
| Refinement method | Full-matrix least-squares on $F^2$ | |
| Data / restraints / parameters | 8481 / 0 / 568 | |
| Goodness-of-fit on $F^2$ | 1.101 | |
| Final R indices [I>2sigma(I)] | R1 = 0.0674, wR2 = 0.1896 | |
| R indices (all data) | R1 = 0.0759, wR2 = 0.1973 | |
| Extinction coefficient | n/a | |
| Largest diff. peak and hole | 0.517 and -1.178 e.$Å^{-3}$ | |

## Table 2. Bond lengths [Å] and angles [°] for *cis*-[$Cd(Tz)_2(py)_2$].

| | | | |
|---|---|---|---|
| Cd(1)-N(1) | 2.218(4) | N(12)-Cd(1)-N(21) | 100.61(16) |
| Cd(1)-N(12) | 2.273(5) | N(1)-Cd(1)-N(11) | 102.29(16) |
| Cd(1)-N(21) | 2.290(4) | N(1)-Cd(1)-N(11) | 99.2(2) |
| Cd(1)-N(11) | 2.313(5) | N(12)-Cd(1)-N(11) | 91.51(17) |
| Cd(1)-N(23) | 2.682(4) | N(21)-Cd(1)-N(11) | 138.21(15) |
| Cd(1)-N(3) | 2.768(4) | N(1)-Cd(1)-N(23) | 132.90(13) |
| N(2)-N(3) | 1.301(5) | N(12)-Cd(1)-N(23) | 92.96(15) |
| N(2)-N(1) | 1.315(5) | N(21)-Cd(1)-N(23) | 50.17(13) |
| N(21)-N(22) | 1.305(6) | N(11)-Cd(1)-N(23) | 89.72(14) |
| N(21)-C(21) | 1.389(6) | N(3)-N(2)-N(1) | 110.4(4) |
| N(22)-N(23) | 1.292(6) | N(23)-N(22)-N(21) | 110.8(4) |

*3.4. Intermolecular interactions landscape*

To achieve a more comprehensive understanding of the intermolecular interactions governing the crystal packing and supramolecular organization of cis-[Cd(Tz)$_2$(py)$_2$], a Hirshfeld surface analysis was carried out using the CrystalExplorer package. The $d_{norm}$ function mapped onto the surface highlights regions of significant intermolecular proximity, where red spots correspond to contacts shorter than the sum of the van der Waals radii, while blue regions indicate comparatively longer separations. In addition to the three-dimensional surface representation, two-dimensional fingerprint plots were employed to resolve and quantify the contributions from distinct atom–atom interactions. This approach enables a detailed and systematic characterization of the full spectrum of intermolecular contacts within the crystal structure.

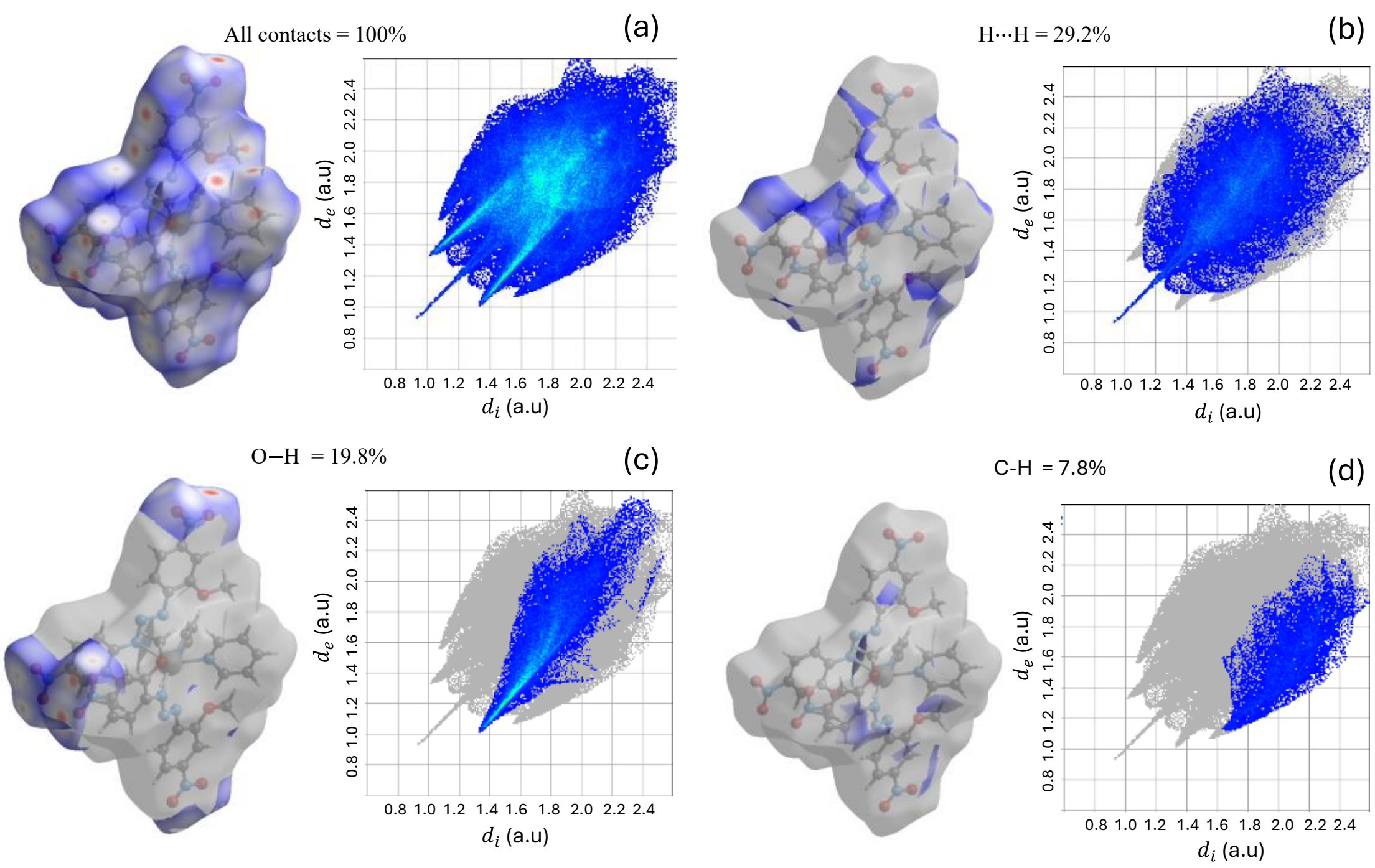

Figure 5: 2D fingerprint plots (overall and decomposed) and the corresponding 3D Hirshfeld surfaces for *cis*-[Cd(Tz)$_2$(py)$_2$] crystal.

Figure 5 presents the overall fingerprint plot together with the decomposed contributions of the dominant intermolecular contacts in cis-[Cd(Tz)$_2$(py)$_2$]. The Hirshfeld surface analysis indicates that the crystal packing is primarily governed by H···H (29.2%), O···H/H···O (19.8%), and C···H/H···C (7.8%) interactions, which together account for 56.8% of the total Hirshfeld surface.

The predominance of H···H contacts reflects the dense packing of hydrogen atoms within the structure, resulting from the compact arrangement of both aromatic and aliphatic moieties. In contrast, the O···H/H···O contacts are indicative of weak C–H···O interactions involving hydrogen atoms from C–H groups and oxygen atoms of the ligand framework, which contribute to the overall stabilization of the supramolecular assembly.

The C···H/H···C contacts, although less significant, are consistent with weak C–H···π interactions. These interactions complement the π–π stacking features identified in the crystal structure (centroid–centroid distance ≈ 3.68 Å), contributing to the organization of the molecular packing.

Focusing on the intermolecular contributions, the shape index surface reveals complementary red–blue triangular patterns characteristic of π–π stacking interactions between aromatic rings, consistent with the intermolecular centroid–centroid distance of approximately 3.68 Å observed in the crystal structure (see Supplementary Figure S3). However, the C···C contacts account for only 3.9% of the Hirshfeld surface in the selected aromatic region, while the overall C···C contribution for the whole molecule is even lower, 2.6% (not shown). This indicates that, although π–π interactions are structurally evident, their contribution to the overall intermolecular packing is relatively modest compared with other contacts.

*3.5. UV-vis absorption analysis*

To investigate the coordination behavior of triazene (Tz) and pyridine (py) ligands toward the $Cd^{2+}$ ion in *cis*-$[Cd(Tz)_2(py)_2]$, the solid-state UV–vis diffuse reflectance spectrum was transformed into the corresponding absorption function, F(R), as depicted in Figure 5, using the Kubelka–Munk formalism. According to this model, F(R) is expressed as [51] :

$$F(R) = (1 - R)^2 / (2R) = K/S \qquad (2)$$

where represents the reflectance (relative relationship between standard and sample $R = R_{sample}/R_{standard}$), $K$ is the absorption coefficient, and $S$ denotes the scattering coefficient. In this framework, $K$ corresponds to the Kubelka–Munk absorption coefficient.

The organometallic complex *cis*-$[Cd(Tz)_2(py)_2]$ (Fig. 5a) exhibits a broad absorption profile extending from approximately 700 nm to 200 nm, indicating a wide range of electronic transitions. In contrast, the free ligand Tz (Fig. 5b) shows absorption starting around 600 nm and extending toward the ultraviolet region. The similarity in the absorption profiles suggests that the electronic transitions in the complex are predominantly ligand-centered, which is consistent with the $d^{10}$ electronic configuration of Cd (II), where metal-to-ligand charge transfer (MLCT) transitions are generally not favored. [24,25,52–54]

The Tauc plot method was employed to estimate the direct and indirect optical band gap energies of the material [55]. This method establishes a relationship between the band gap energy and the absorption coefficient ($\alpha$), based on the Tauc equation: $\alpha h\nu = A(h\nu - E_g)^m$, where $h$ is Planck's constant, $\nu$ is the radiation frequency, $A$ is a constant, and $m$ is associated with the nature of the electronic transition. For allowed direct transitions, $m = 1/2$, while for allowed indirect transitions, $m = 2$. Considering that the Kubelka–Munk function, $F(R)$, is proportional to the material's absorption coefficient (only at the absorption edge, where the reflectance is approximately constant), the Tauc relation can be rewritten as $F(R)h\nu = A(h\nu - E_g)^m$.

Tauc plots derived from the absorption data (Fig. 6) indicate direct bandgap behavior for both samples. The estimated optical bandgap values are 1.83 eV for *cis*-$[Cd(Tz)_2(py)_2]$ and 2.14 eV for the free ligand Tz. In addition, the indirect bandgap values were determined to be 1.73 eV for *cis*-$[Cd(Tz)_2(py)_2]$ and 2.01 eV for the free ligand (see Fig. S4). Interestingly, the estimated direct band gap is significantly lower than those reported for related Cd(II)-based complexes, suggesting differences in the electronic structure arising from the ligand environment and coordination effects. [23,56]

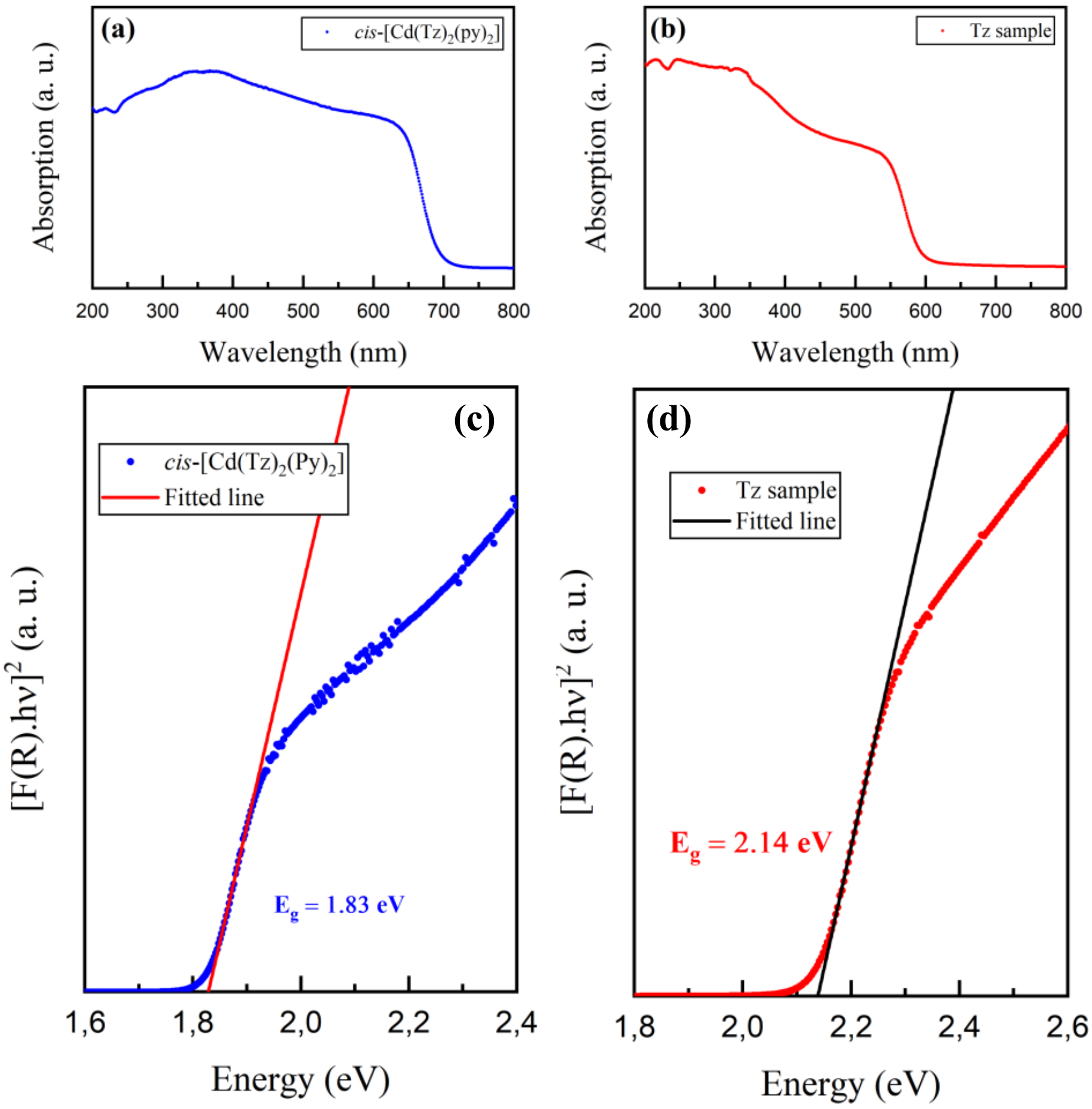

Figure 5. Absorption spectra for (a) *cis*-[Cd(Tz)$_2$(py)$_2$] and (b) Tz sample. Tauc plots considering direct (c, d) gaps, respectively.

*3.6. Photoluminescence properties*

Taking into account the excellent luminescent properties of the π conjugation species and $d^{10}$ metal ions compounds, the luminescence properties of the free Tz ligand and the *cis*-[Cd(Tz)$_2$(py)$_2$] were investigated in the solid state at room temperature.

The emission spectra of the *cis*-[Cd(Tz)$_2$(py)$_2$] complex and the Tz ligand are shown in Figure 7a. The complex exhibits a broad emission band in the 500–850 nm region, with two main contributions centered around 575 and 700 nm ($\lambda_{ex}$ = 457 nm). In contrast, the free ligand displays a more intense emission centered at approximately 600 nm ($\lambda_{ex}$ = 457 nm), with a less intense contribution in the red region (~700 nm). Upon coordination to Cd(II) and pyridine, the emission profile becomes broader and shows enhanced intensity in the red region.

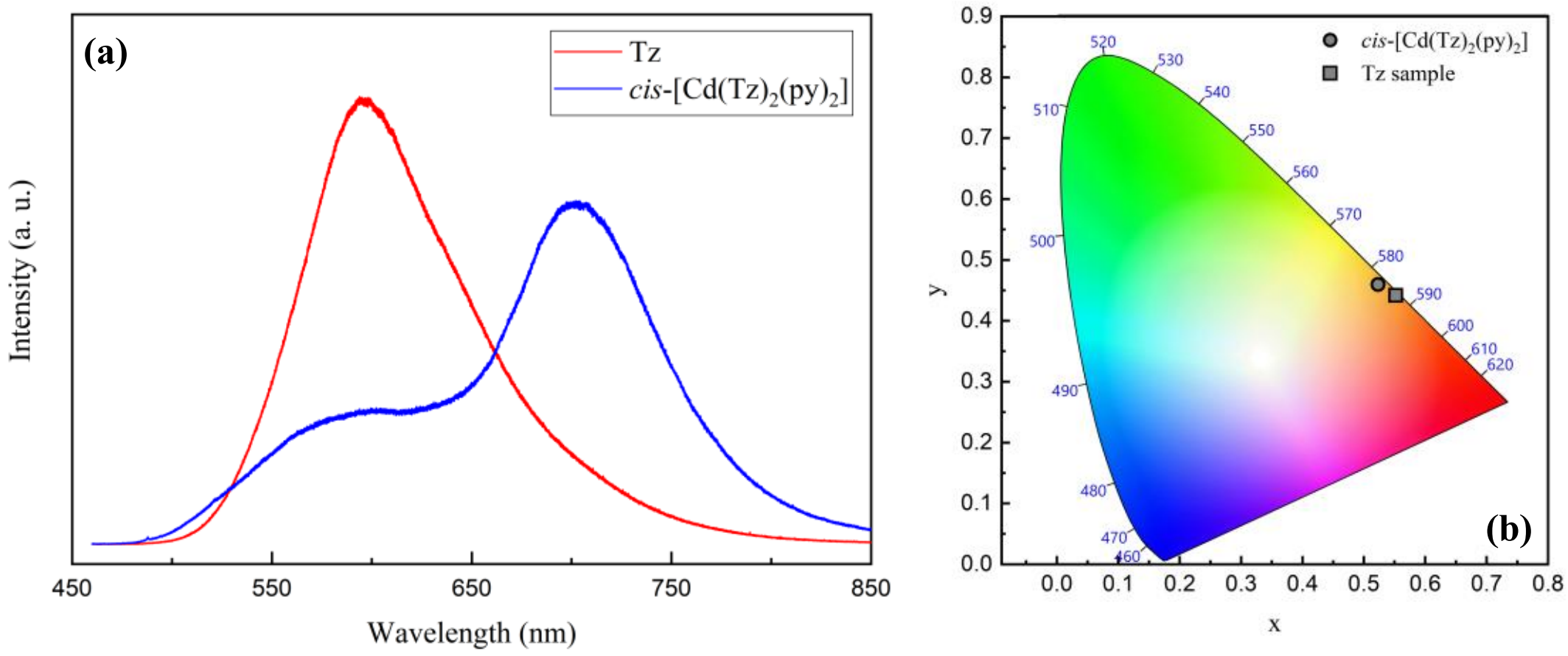

Figure 7. (a) PL emission spectrum of organic Tz and complex *cis*-$[Cd(Tz)_2(py)_2]$ at 300 K. (b) Shows the colour chromaticity diagram (CIE).

In order to further correlate these spectral features with the perceived emission color, the CIE 1931 chromaticity coordinates were determined from the emission data presented in Figure 7b. The Tz ligand exhibits coordinates of (0.5521, 0.4417), while the *cis*-$[Cd(Tz)_2(py)_2]$ complex shows coordinates of (0.5229, 0.4597). These values place both emissions in the warm region of the chromaticity diagram, consistent with the dominant yellow–orange to red contributions observed in their emission spectra. The corresponding correlated color temperatures (CCTs) were estimated to be approximately 1974 K for Tz and 2325 K for the complex, indicating a slight shift in the emission profile upon coordination, while preserving the overall warm character. These features highlight its potential for application in red-emitting light-emitting diodes.

Although the absorption spectra suggest predominantly ligand-centered transitions, the photoluminescence behavior is significantly altered upon complexation. This effect can be attributed to changes in the excited-state dynamics induced by coordination, such as increased structural rigidity and modification of non-radiative decay pathways. In this context, the X-ray structure reveals that the bidentate coordination of the triazenide ligands and the distorted octahedral geometry around Cd(II) can perturb the electronic distribution within the π-conjugated framework, thereby modifying the energy of the ligand-centered excited states. In agreement with literature reports for Cd(II) complexes, the $d^{10}$ electronic configuration prevents metal-centered and charge-transfer (MLCT/LMCT) transitions, indicating that the observed luminescence arises mainly from ligand-centered (LC) transitions (π→π* and n→π*). [24,25,54]

The changes in emission intensity and band profile may be associated with energy transfer processes and enhanced non-radiative relaxation pathways, which can lead to fluorescence quenching or redistribution of emission intensity. Additionally, the slight contraction of the N–N–N angle upon coordination and the preservation of N–N bond distances suggest subtle modifications in electron delocalization within the triazene unit, which may further contribute to the observed spectral shifts.

The Voigt fitting analysis further supports the presence of multiple emissive processes. The Tz ligand shows three emission components (Fig. S5), whereas the complex exhibits four (Fig. S6), suggesting that coordination introduces additional emissive pathways. Notably, the emission at ~700 nm becomes more prominent in the complex, indicating a redistribution of excited-state population upon coordination. [21,22,24] Moreover, the intra- and intermolecular π–π stacking interactions observed in the crystal packing may play a secondary role in influencing excited-state relaxation processes in the solid state. However, the photoluminescence behavior is predominantly governed by ligand-centered (LC) transitions modulated by coordination to the Cd(II) center.

## 4.Conclusions

In summary, a new organometallic complex, *cis*-$[Cd(Tz)_2(py)_2]$, featuring a distorted octahedral Cd(II) coordination geometry, was successfully synthesized and structurally characterized. The Cd(II) center is coordinated by two bidentate triazenide ligands and two pyridine molecules, as confirmed by X-ray crystallographic analysis, revealing subtle structural modifications such as N–N–N angle contraction and the presence of intra- and intermolecular π–π interactions. The electronic properties are primarily governed by ligand-centered (LC) transitions modulated by coordination to the Cd(II) center, whereas intermolecular interactions contribute to the overall crystal packing.

The crystal packing is governed by a combination of intermolecular interactions, predominantly H···H (29.2%), O···H/H···O (19.8%), and C···H/H···C (7.8%) contacts, which together dominate the Hirshfeld surface and reflect the compact molecular arrangement. The O···H/H···O and C···H/H···C contributions are associated with weak C–H···O and C–H···*π* interactions that contribute to the stabilization of the structure. In contrast, both intra- and intermolecular π–π stacking interactions, although structurally evident and supported by shape index analysis, make only a modest contribution to the overall packing.

Diffuse reflectance UV–Vis spectroscopy indicates a direct optical band gap of 1.83 eV for the complex, significantly lower than that of the free ligand, suggesting coordination-induced modification of the electronic structure and behavior consistent with semiconductor-like materials. The absorption profile, largely preserved upon coordination, supports the predominance of ligand-centered (LC) transitions, consistent with the $d^{10}$ electronic configuration of Cd(II).

Photoluminescence studies reveal a broad emission band (500–850 nm) with enhanced red emission upon coordination, attributed to ligand-centered (π→π* and n→π*) transitions. The observed changes in emission profile, including band broadening and increased red contribution, are associated with modifications in excited-state dynamics, energy transfer processes, and non-radiative relaxation pathways induced by coordination. Voigt fitting analysis further confirms the presence of multiple emissive components, indicating increased complexity of the excited-state landscape in the complex. The CIE chromaticity coordinates confirm warm emission characteristics, highlighting the potential of this system for application in red-emitting light-emitting diodes and other optoelectronic devices.

## Acknowledgments

Ariel Nonato acknowledges financial support from the Talent4Iberia Project (Grant No. 101128265, MSCA-COFUND 2022), funded under the Horizon Europe Programme. This work was partially supported by the Spanish Ministry of Science, Innovation and Universities and the Regional Government of Extremadura through the CIIAE Unit. We acknowledge the support of the INCT project Advanced Quantum Materials, involving the Brazilian agencies CNPq (Proc. 408766/2024-7), FAPESP (Proc. 2025/27091-3), and CAPES.

**Supplementary material**

# Coordination-Induced Tuning of Ligand-Centered Red Emission in a *cis*-[Cd(Tz)$_2$(py)$_2$] Complex for Light-Emitting Diodes

Samara M. da Silva[1‡], R. F. Silva[6], A. Nonato[2,3‡*], Paulo Villis[4], Rodrigo S. Corrêa[5], L. C. Gómez-Aguirre[2], C.W.A. Paschoal[6], Pedro I. S. Maia[7], Benedicto A. V. Lima[8,*]

[1] *Coordenação de Ciências Naturais - Química, Universidade Federal do Maranhão, Centro de Ciências de Grajaú- CCGR, 65.940-000, Grajaú, MA, Brazil.*

[2] *Department of Thermal Energy Storage, Iberian Centre for Research in Energy Storage CIIAE, 10003 Caceres, Spain*

[3] *Programa de Pós-graduação em Física, Universidade Federal do Maranhão, Campus Universitário do Bacanga, São Luís, Maranhão 65080-805, Brazil*

[4] *Laboratory of Electrochemistry and Biotechnology, Ceuma University, São Luís, postal code 65.065-470, MA, Brazil.*

[5] *Departamento de Química, Instituto de Ciências Exatas e Biológicas (ICEB), Federal University of Ouro Preto, Morro do Cruzeiro Campus, postal code 35.400-000, Ouro Preto, MG, Brazil.*

[6] *Departamento de Física, Universidade Federal do Ceará, Campus do Pici, 65455-900, Fortaleza - CE, Brazil*

[6] *Departamento de Química, Instituto de Ciências Exatas e Biológicas (ICEB), Universidade Federal de Ouro Preto, 35400-000 Ouro Preto-MG, Brazil*

[7] *Núcleo de Desenvolvimento de Compostos Bioativos (NDCBio), Universidade Federal do Triângulo Mineiro, Av. Dr. Randolfo Borges 1400, Uberaba, Minas Gerais, Brazil*

[8] *Medicinal Chemistry and Biotechnology Research Group, Coordenação de Ciências Naturais - Química, Universidade Federal do Maranhão, Centro de Ciências de São Bernardo - CCSB, 65550-000, São Bernardo, MA, Brazil*

[*] Corresponding author. Tel: +34 610281478

E-mail address: ariel.almeida@ciiae.org;

‡ These authors contributed equally to this work.

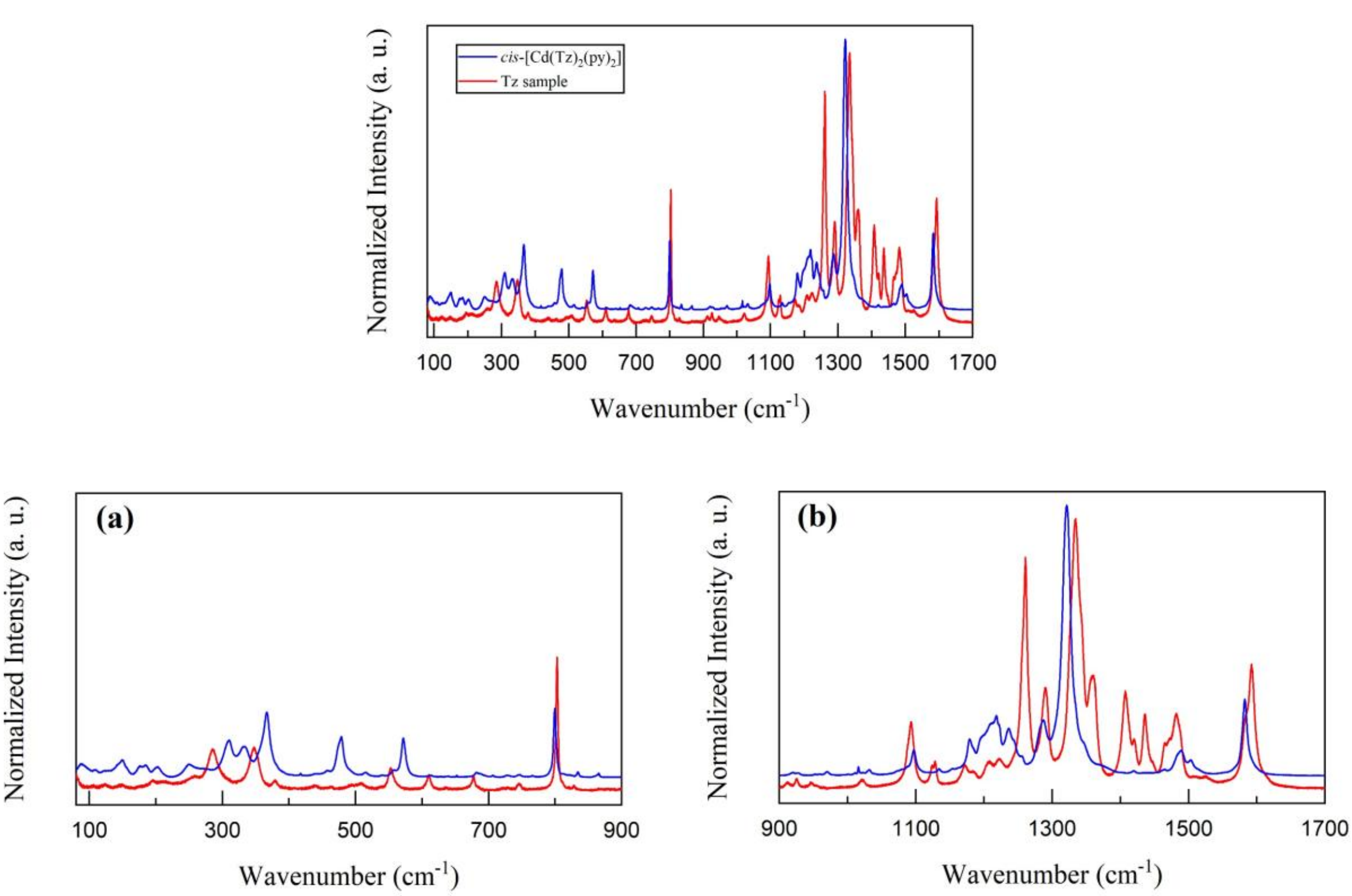

Fig. S1: Normalized vibrational spectra (IR/Raman) of the free ligand (Tz, red) and the complex cis-[$Cd(Tz)_2(py)_2$] (blue). The top panel shows the full spectral range (100–1700 $cm^{-1}$), while panels (a) and (b) present expanded views of the low-frequency (100–900 $cm^{-1}$) and high-frequency (900–1700 $cm^{-1}$) regions, respectively. All spectra are presented in normalized form within the 0–1 range.

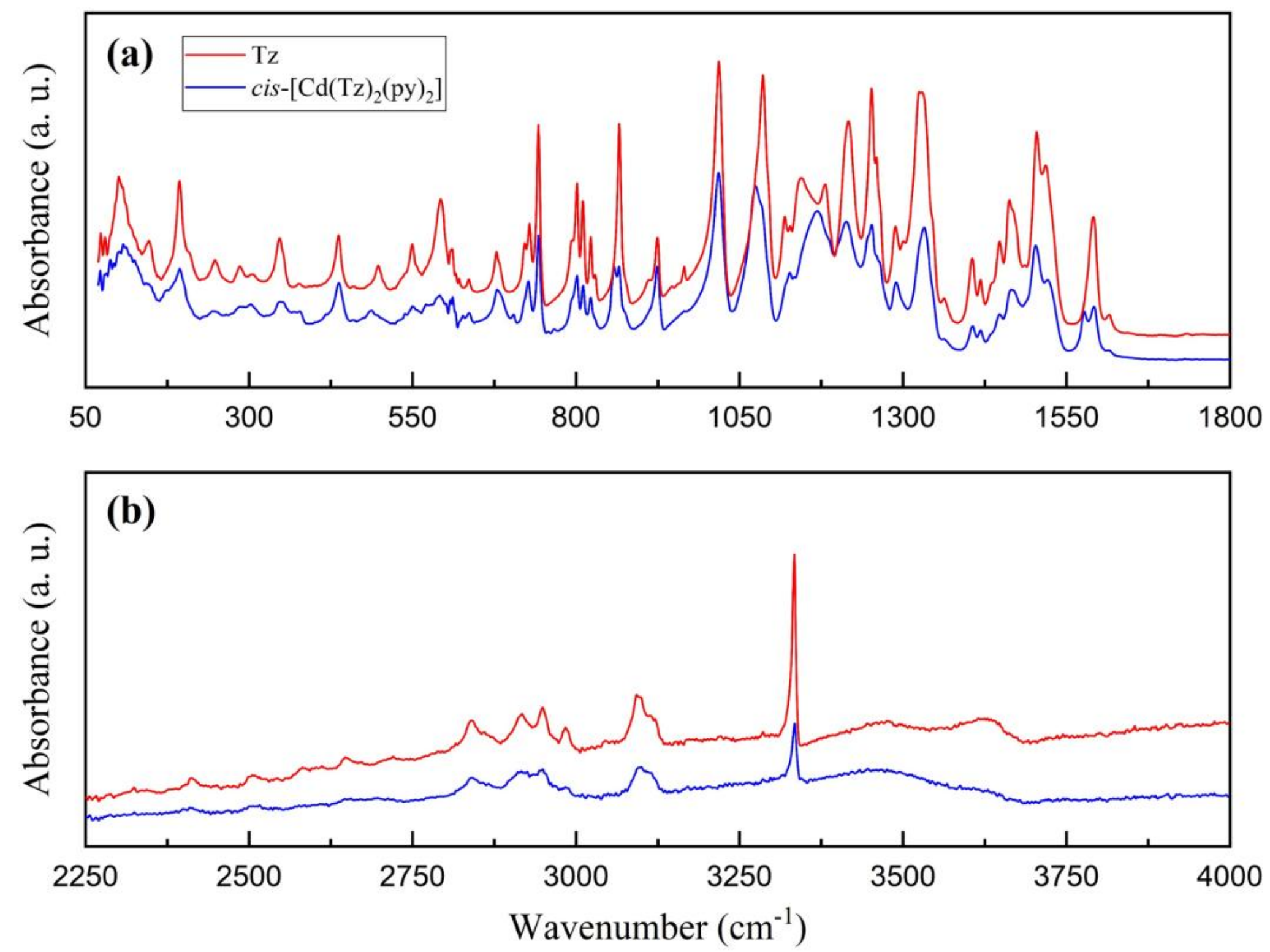

Fig. S2: FTIR-ATR spectra of the free ligand (Tz, red) and the complex cis-$[Cd(Tz)_2(py)_2]$ (blue). Panel (a) shows the fingerprint region (50–1800 $cm^{-1}$), while panel (b) displays the high-wavenumber region (2250–4000 $cm^{-1}$). All spectra are presented in normalized form within the 0–1 range.

**Table 1**: Observed frequencies and assignment of some Raman and IR modes for *cis*-$[Cd(Tz)_2(py)_2]$

| Mode | Simmetry | Assignment [1–7] | Raman | | Infrared | |
|---|---|---|---|---|---|---|
| | | | **Frequency ($cm^{-1}$)** | **Intensity** | **Frequency ($cm^{-1}$)** | **Intensity** |
| | | | 87 | m | - | - |
| | | | 95 | m | - | - |
| | | | 108 | w | - | - |
| | | | 110 | w | 110 | s |
| | | | 122 | w | - | - |
| | | | 129 | w | - | - |

| | | | | | |
|---|---|---|---|---|---|
| | | 143 | m | - | - |
| | | 150 | m | 148 | v-w |
| | | 157 | w | - | - |
| | | 174 | m | | |
| | | 178 | m | 177 | w |
| | | 185 | m | - | - |
| | | - | - | 194 | w |
| | | 202 | m | - | - |
| | | - | - | 246 | w |
| | | 249 | m | - | - |
| | | 262 | w | - | - |
| | | - | - | 265 | v-w |
| | | 275 | m | 272 | v-w |
| | | - | - | 283 | w |
| | | | | 292 | v-w |
| | | | | 302 | w |
| | | 308 | s | - | - |
| | | 311 | m | 312 | w |
| | | - | - | 327 | v-w |
| | | 330 | m | - | - |
| E´´ | Out-of-plane ring bending | 336 | m | - | - |
| | | - | - | 349 | w |
| | | 357 | m | - | - |
| | | 367 | v-s | - | - |
| | | - | - | 372 | v-w |
| | | 381 | v-w | 379 | v-w |
| E´´ | Quadrant out-of-plane bend, ring | 417 | v-w | 418 | w |
| | | 441 | v-w | 438 | m |
| | | 457 | w | 459 | v-w |
| | | - | - | 471 | v-w |
| | | 474 | s | - | - |
| | | 479 | s | - | - |
| | | - | - | 486 | w |
| | | 490 | w | - | - |
| | | - | - | 497 | v-w |
| | | - | - | 507 | w |
| | | 515 | w | - | - |
| | | - | - | 529 | v-w |
| | | 540 | v-w | 538 | v-w |
| | | - | - | 551 | w |
| | | 556 | w | - | - |
| | | 559 | w | - | - |
| | | 572 | s | 571 | w |
| | | 575 | m | - | - |

| | | | | | | |
|---|---|---|---|---|---|---|
| | | | - | - | 578 | v-w |
| | | | - | - | 586 | w |
| | | | - | - | 592 | w |
| | | | - | - | 597 | w |
| | | | - | - | 601 | v-w |
| | | | - | - | 607 | w |
| | | | 612 | v-w | 612 | w |
| | | | - | - | 616 | w |
| | | | - | - | 622 | v-w |
| | | | 628 | v-w | 627 | v-w |
| | | | 638 | v-w | 636 | w |
| | | N−C axial rocking + all $CH_2$ rocking | 652 | v-w | - | - |
| | | | - | - | 658 | v-w |
| | E´ | Ring vibration | - | - | 679 | m |
| | | | 682 | w | - | - |
| $\nu_{10}$ | E´ | Quadrant in-plane bend, ring | 688 | w | 688 | w |
| | | | 693 | v-w | - | - |
| | | | - | - | 704 | v-w |
| | | | 707 | v-w | - | - |
| | | | 728 | v-w | 726 | m |
| | | | 731 | v-w | - | - |

| $A''_2$ | Ring out-of-plane bend + CH wag (in-phase) | 742 | v-w | 743 | s |
|---|---|---|---|---|---|
| | | 746 | v-w | - | - |
| | | - | - | 758 | v-w |
| | | 766 | v-w | 767 | v-w |
| | | - | - | 778 | v-w |
| | | 799 | s | 794 | m |
| | | 802 | w | 801 | m |
| | | 812 | v-w | 811 | m |
| | | 816 | v-w | - | - |
| | | 822 | v-w | 822 | m |
| | | 826 | v-w | 827 | w |
| | | 834 | w | - | - |
| | | 835 | v-w | - | - |
| | | - | - | 842 | w |
| | | 860 | v-w | 858 | m |
| | | 863 | v-w | - | - |
| | | 866 | w | 866 | m |
| | | - | - | 868 | w |
| | | - | - | 876 | w |
| | | - | - | 901 | w |
| | | - | - | 914 | w |

| | | | | | | |
|---|---|---|---|---|---|---|
| | | | 918 | v-w | - | - |
| | | | 920 | v-w | 922 | m |
| | | | 926 | v-w | 926 | m |
| | | | 928 | v-w | - | - |
| | | | 952 | v-w | - | - |
| $\nu_5$ | $A'_2$ | Sextant stretch, ring | 958 | v-w | 963 | w |
| | | | 967 | v-w | - | - |
| | | | 970 | v-w | - | - |
| $\nu_3$ | $A'_1$ | N radial, in-phase | 972 | v-w | - | - |
| | | | - | - | 994 | w |
| | | | 1000 | v-w | - | - |
| | | | 1008 | v-w | - | - |
| | | | 1011 | v-w | 1011 | m |
| | | | 1016 | w | 1019 | v-s |
| | | | 1021 | v-w | - | |
| | | | 1029 | v-w | - | - |
| | | Asymmetric N−C stretching | 1030 | w | - | - |
| $\nu_{12}$ | $E''$ | CH wag, out-of-phase | 1031 | w | - | - |
| | | | 1034 | v-w | - | - |
| | | | 1038 | v-w | - | - |
| | | | 1041 | v-w | - | - |
| | | | 1061 | v-w | - | - |
| | | | 1072 | v-w | 1073 | v-s |
|  |  |  | 1079 | w | - | - |
|  |  |  | 1087 | v-w | 1086 | m |
|  |  |  | 1097 | s | - | - |
|  |  |  | - | - | 1120 | w |
|  |  |  | - | - | 1127 | m |
| $\nu_2$ | $A'_1$ | C radial, in-phase | 1133 | w | - | - |
| $\nu_9$ | $E'$ | Semi-circle str, rign + CH rk | 1153 | w | 1148 | m |
|  |  |  | 1162 | w | 1166 | s |
|  | $E'$ | CH rk | 1178 | s | 1180 | s |
|  |  |  | 1184 | w | - | - |
|  |  |  | 1194 | s | - | - |
|  |  |  | 1200 | s | - | - |
|  |  |  | 1209 | s | - | - |
|  |  |  | 1218 | s | 1214 | s |
|  |  |  | 1223 | s | - | - |
|  |  |  | 1236 | s | - | - |
|  |  |  | 1244 | s | 1248 | s |
|  |  |  | 1252 | w | 1253 | s |
|  |  |  | 1256 | w | - | - |
|  |  |  | - | - | 1264 | m |
|  |  |  | 1280 | s | - | - |
|  |  |  | 1287 | s | 1291 | m |
|  |  |  | 1303 | w | - | - |
|  |  | Ring’s C – C stretching vibrations | 1321 | v-s | 1326 | s |
| | | | 1335 | w | 1335 | s |
| $\nu_4$ | $A'_2$ | CH rk, in phase | 1346 | s | 1344 | w |
| | | | 1373 | w | - | - |
| | | | - | - | 1406 | w |
| $\nu_8$ | $E'$ | CH rk + ring semi-circle str | 1419 | w | 1419 | w |
| | | | - | - | 1435 | v-w |
| | | | - | - | 1447 | w |
| | | | 1461 | v-w | - | - |
| | $E'$ | Ring semi-circle str | 1465 | v-w | 1463 | m |
| | | | - | - | 1472 | m |
| | | Ring's C – C stretching vibrations | 1483 | s | 1484 | w |
| | | | 1490 | s | - | - |
| | | | 1503 | s | 1503 | s |
| | | Asymmetric – $NO_2$ stretching | 1514 | v-w | - | - |
| | | | - | - | 1521 | s |
| | | | - | - | 1530 | w |
| | $E'$ | Ring vibration | 1567 | v-w | - | - |
| $\nu_7$ | $E'$ | Quadrant str, ring + CH rk | 1582 | v-s | 1577 | m |
| | | | - | - | - | - |
| | | | - | - | 1592 | m |
| | | | - | - | 1616 | v-w |
| | | | - | - | 2411 | |
| | | | - | - | 2508 | |

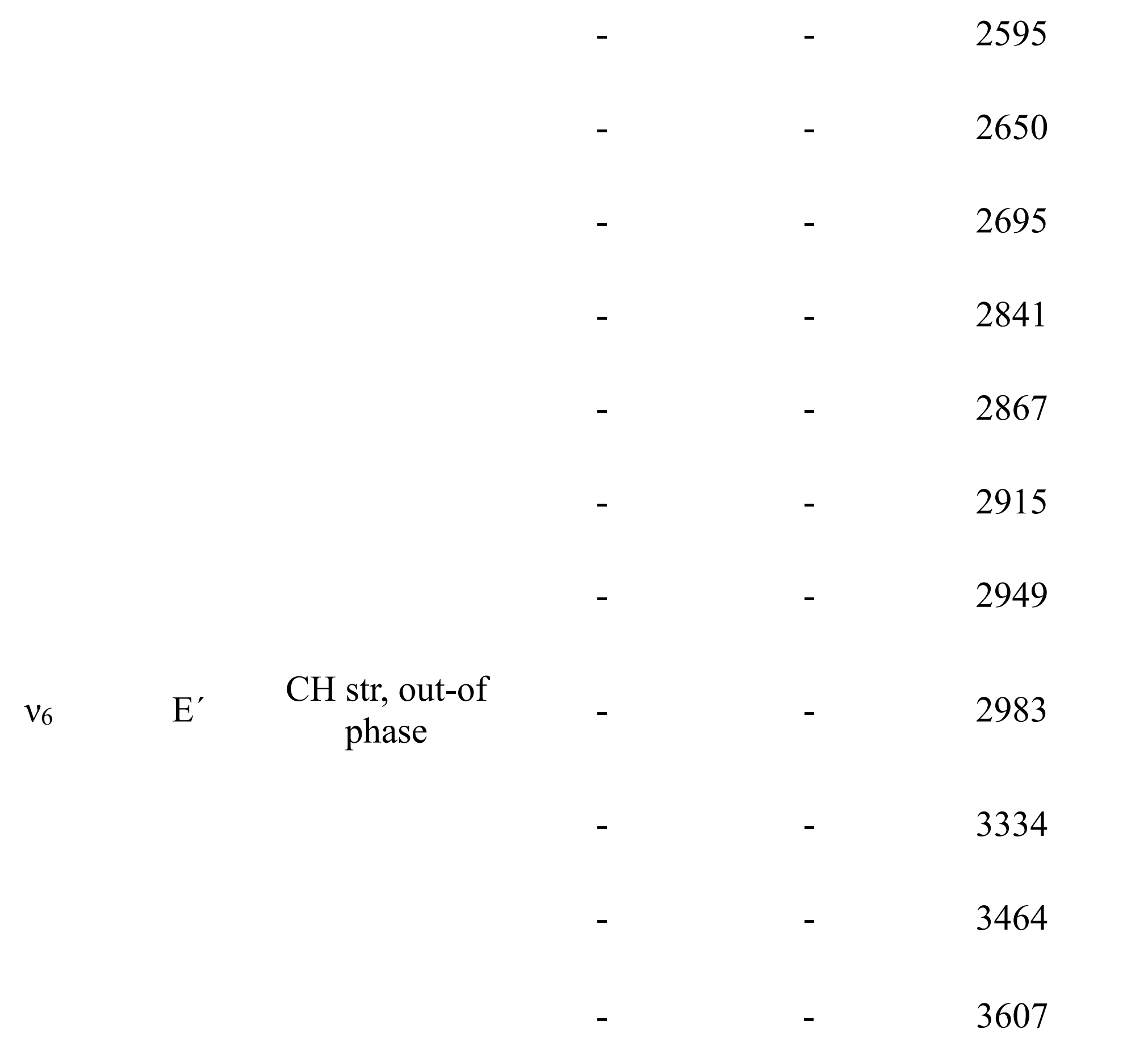

| | | | | | |
|---|---|---|---|---|---|
| | | | - | - | 2595 |
| | | | - | - | 2650 |
| | | | - | - | 2695 |
| | | | - | - | 2841 |
| | | | - | - | 2867 |
| | | | - | - | 2915 |
| | | | - | - | 2949 |
| $\nu_6$ | E′ | CH str, out-of phase | - | - | 2983 |
| | | | - | - | 3334 |
| | | | - | - | 3464 |
| | | | - | - | 3607 |

Key: s-very strong, s-strong, m-medium, w-weak, vw-very weak;

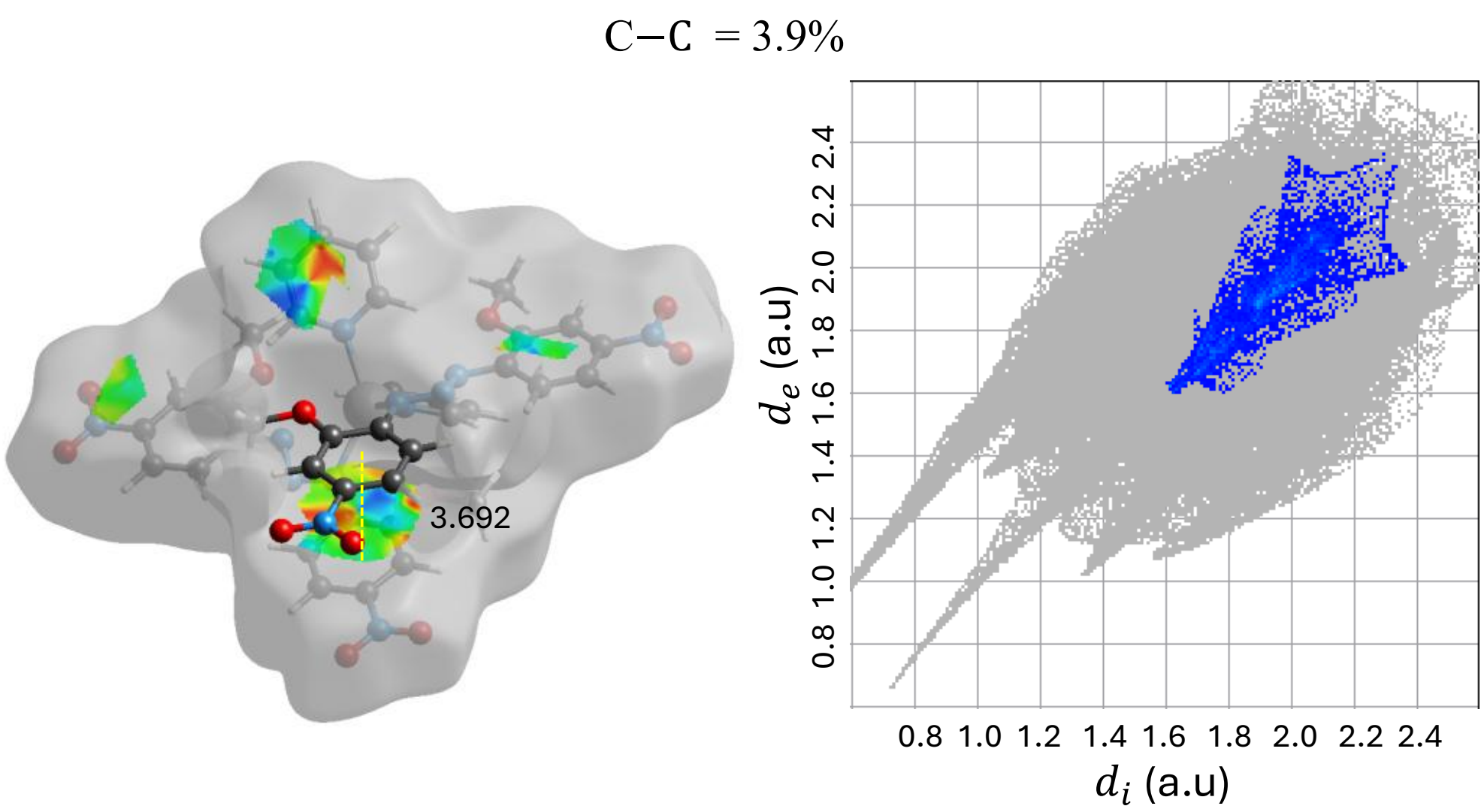

Fig. S3. Hirshfeld surface mapped over $d_{norm}$ (left) and the corresponding two-dimensional fingerprint plot highlighting C···C contacts (right). The Hirshfeld surface was additionally analyzed using the shape index function to emphasize the π–π stacking features identified in the crystal structure, showing complementary red–blue triangular patterns between adjacent aromatic rings.

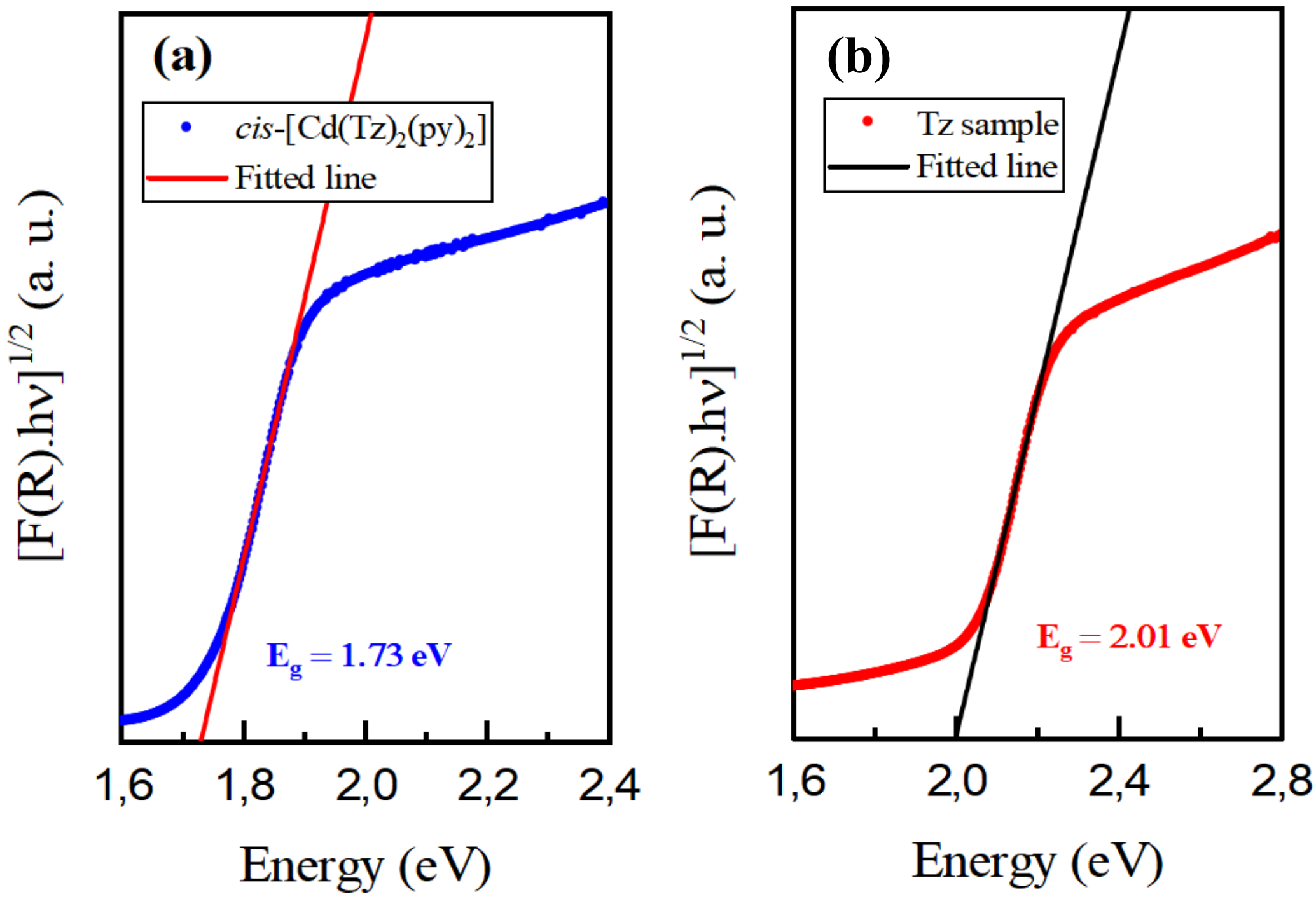

Fig. S4: Tauc plots considering indirect gap.

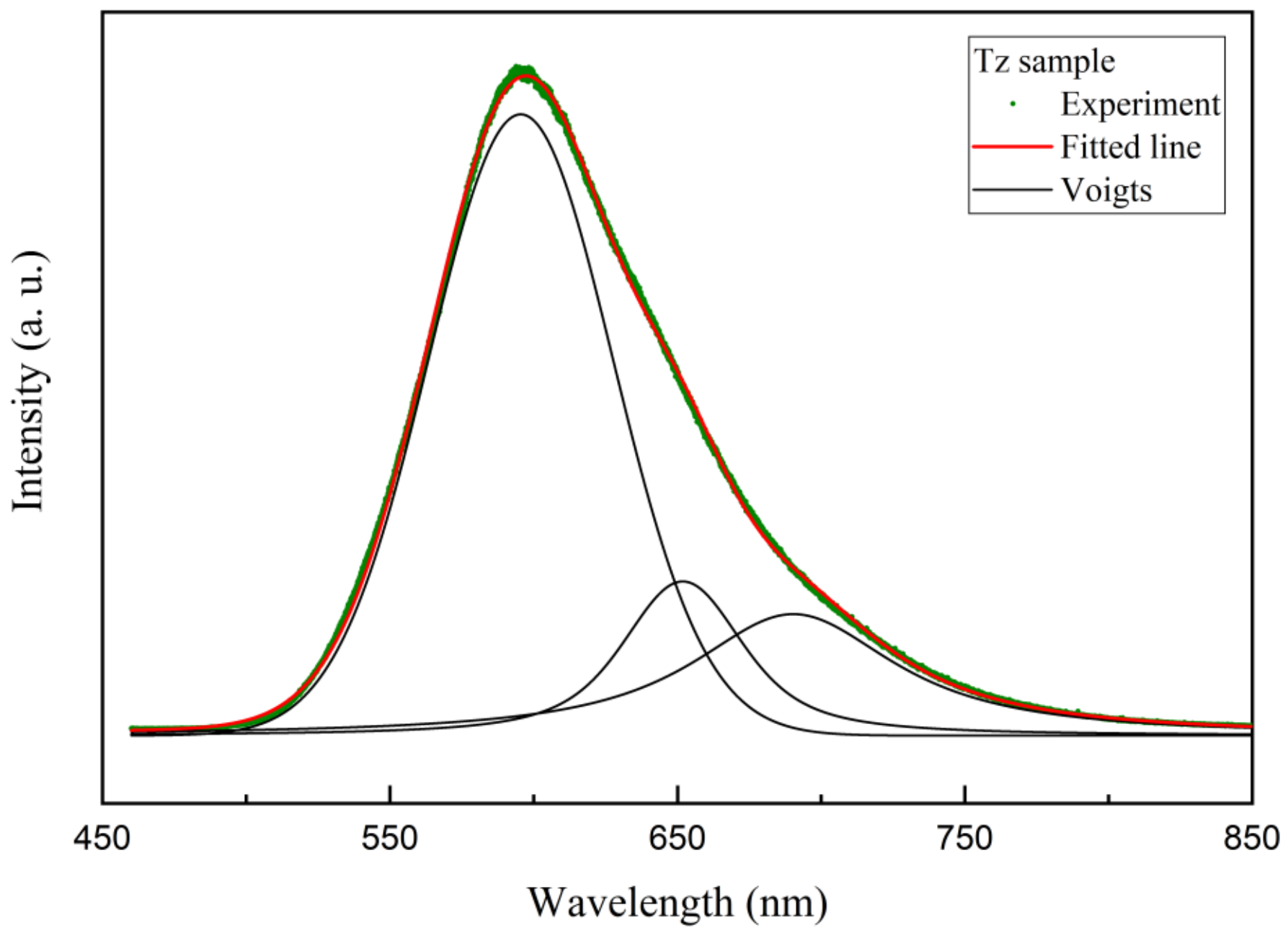

Fig. S5: Deconvolution of the PL emission spectrum of organic Tz.

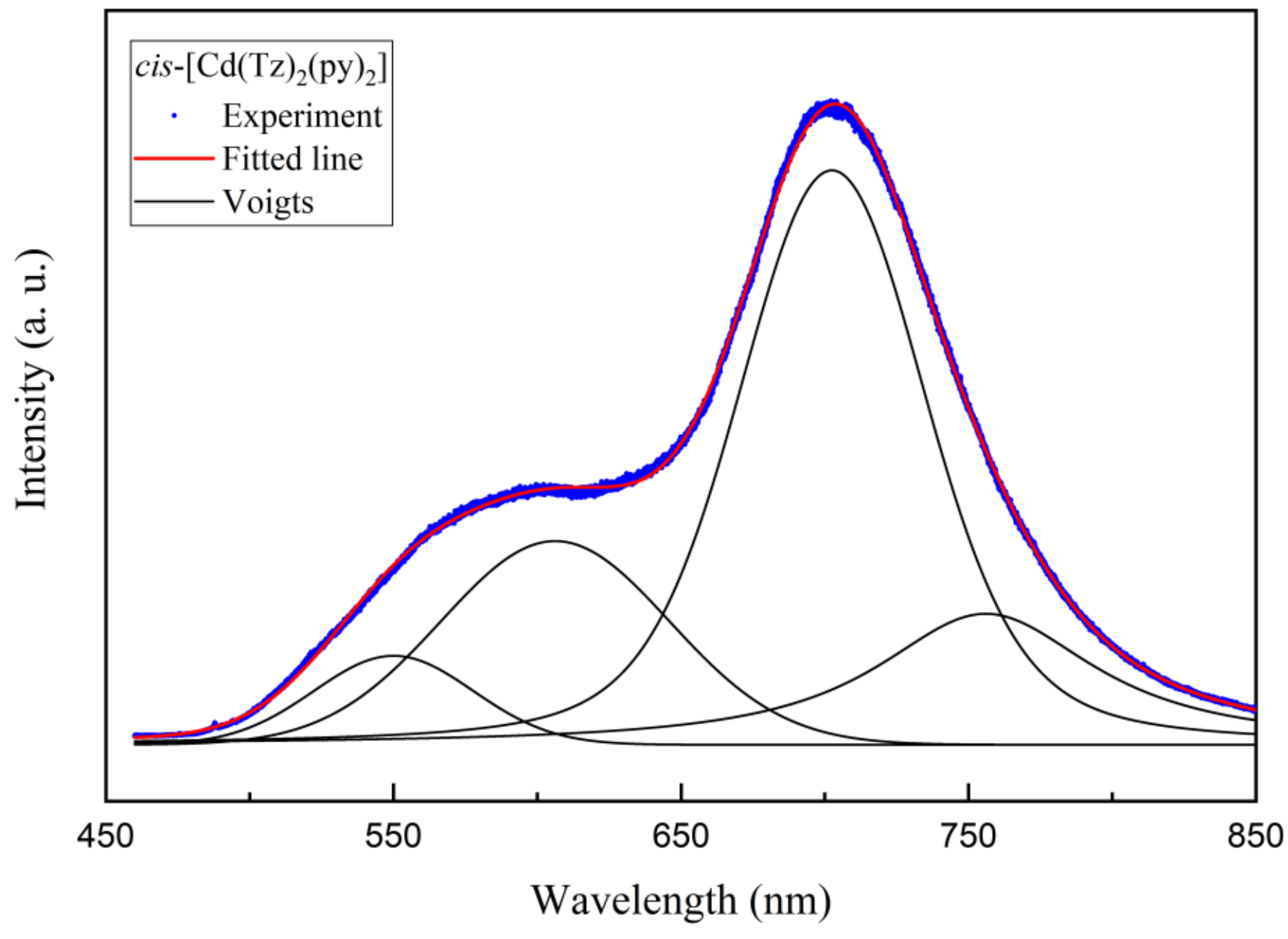

Fig. S6: Deconvolution of the PL emission spectrum of cis-$[Cd(Tz)_2(py)_2]$ complex.